\documentclass[10pt,leqno]{amsart}
\usepackage{graphicx}
\usepackage{indentfirst,csquotes}
\usepackage{booktabs,tabularx,array}

\usepackage{amssymb,amsthm,amsmath}
\usepackage{xcolor,paralist,fancyhdr,etoolbox}

\usepackage{lipsum}

\makeatletter
\newcommand{\affiliations}[1]{\gdef\@affiliations{#1}}
\let\@affiliations\@empty
\apptocmd{\@setauthors}{%
  \ifx\@affiliations\@empty\else
    \par\vspace{0.5\baselineskip}%
    {\centering\footnotesize\itshape \@affiliations\par}%
  \fi
}{}{\PackageWarning{affil}{Failed to patch \string\@setauthors}}
\makeatother

\usepackage{natbib}
\usepackage{hyperref} 

\hypersetup{ colorlinks=true, linkcolor=black, filecolor=black, urlcolor=black }

\usepackage{lipsum}

\begin{document}
\title[Chinese-Language Coordinated Activity in Japanese-Language X]{Crude, Commercial, and Self-Referential: Chinese-Language Coordinated Activity in Japanese-Language X}


\author{Kei Ichikawa\textsuperscript{$\dagger$,*}}
\author{Bruno T. Sugano\textsuperscript{$\dagger$}}
\author{Genta Toya\textsuperscript{$\dagger$}}
\author{WU QIANYUN\textsuperscript{$\ddagger$}}
\author{Yasuhiro Hashimoto\textsuperscript{$\S$}}
\author{Masashi Toyoda\textsuperscript{$\P$}}
\author{Naoki Yoshinaga\textsuperscript{$\P$}}
\author{Kazutoshi Sasahara\textsuperscript{$\dagger$}}

\affiliations{%
  \textsuperscript{$\dagger$}Institute of Science Tokyo\\
  \textsuperscript{$\ddagger$}Hirosaki University\\
  \textsuperscript{$\S$}University of Aizu\\[3pt]
  \textsuperscript{$\P$}The University of Tokyo\\
  \textsuperscript{*}Corresponding author: \href{mailto:ichikawa.k.25a5@m.isct.ac.jp}{\upshape ichikawa.k.25a5@m.isct.ac.jp}%
}

\makeatletter
\def\shortauthors{Ichikawa et al.}
\makeatother


\begin{abstract}
Malicious coordination has long been regarded as a principal source of information ecosystem pollution. Here, we focus on crude, text-repetition-based coordination. As the demand for mitigating its dissemination has grown, scholars have studied such coordination, focusing especially on bot detection. Few studies, however, have characterized malicious coordination per se or examined how it elicits reactions from general users. Leveraging a dataset of 734,173 Chinese-language coordinated accounts and around 495 million coordinated posts published between May 2024 and March 2026, this study analyzes the characteristics of coordinated behavior and how general users react to coordinated posts. We report three findings: (1) most coordinated accounts are crude and retain the classic marks of automation, and the same criterion applied to Japanese-language accounts over the same month yields a share six times lower; (2) their content is overwhelmingly non-political; (3) regarding their reach, most reactions within large observable cascades originate from coordinated accounts themselves, while posts classified as potentially harmful or illegal material receive a comparatively high proportion of reactions from outside the Chinese-dominant population. We provide a longitudinal quantitative map of crude Chinese-language coordination appearing in X’s Japanese-classified stream.
\end{abstract} 
\maketitle
\renewcommand{\shortauthors}{Ichikawa et al.}

\bigskip


$\,$
   \section{Introduction}
    
    The pollution of information ecosystems has long been a central concern of social media research \citep{pacheco2021uncovering, lazer2018science}. Malicious coordination---regardless of its automated or organic nature---has been regarded as one of the principal drivers of this degradation \citep{mannocci2024detection, pacheco2021uncovering}. Because coordinated accounts can inject a large volume of messages into circulation at once, they can raise users' exposure to particular content \citep{stella2018bots, shao2018spread, torres2022manufacture} and distort what users take the prevailing opinion to be \citep{ross2019social}. Coordination can also work by hindering search behavior: an information ecosystem saturated with noise is one in which anything, true or false, becomes harder to find \citep{Roberts+2018+190+222}. Considerable attention has therefore been paid to coordination and to preventing its dissemination.

    To prevent the spread of malicious coordination, most research to date has focused on detecting social bots \cite{ferrara2016rise, varol2017online}. Because most researchers use supervised machine learning models for detection, much of their attention has been devoted to creating labeled datasets \cite{pacheco2021uncovering}. In addition to these approaches, some scholars have argued that supervised machine learning models are not suited to detecting coordinated accounts, because they rely mainly on features drawn from individual accounts or tweets \cite{chen2018unsupervised, cresci2017paradigm, grimme2018changing}. While detection is a highly important topic, analyzing every account in order to identify coordinated ones is inefficient. If we knew which types of coordinated accounts are more influential, we could prioritize those for detection and for mitigating their dissemination. The same holds for what they post. If coordination is assumed to follow specific events e.g., political events, detection effort will be concentrated around elections and crises. However, their actual reach and factors associated with their activity remain poorly understood. In this study, we focus specifically on crude coordination manifested through repeated identical or near-identical text across multiple accounts, rather than attempting to capture coordination in all its forms.

    Few studies have addressed how coordination influences user behavior \citep{stella2018bots, bail2020assessing}. For example, \citet{bail2020assessing} find that the Russian campaign directed at the United States in 2017 did not affect political attitudes. A number of studies have examined coordination in the context of political events; however, coordinated accounts may be active not only around political topics but also around other topics. In addition, event-centered studies often examine political events one at a time, so the characteristics of coordination may look different when observed over a long span rather than at a single point.
    
    Leveraging a data set of 734,173 Chinese-language coordinated accounts and 495 million coordinated posts from X between May 2024 and March 2026, we analyze how this form of coordination unfolds over roughly two years. We ask three things: (i) what characterizes coordinated accounts, (ii) whether surges in their activity are organized around political events, and (iii) whether their posting elicits any reactions (replies, reposts, quote posts) from accounts outside the coordinated population.

    \section{Related Work}
    
    \subsection{Detecting Coordinated Behavior}
    Research on coordinated behavior in social media has proceeded mainly by focusing on three things: the objects shared between accounts (URLs, hashtags and the like), behavioral patterns (e.g., posting, reposting), and the times at which accounts are active \cite{mannocci2024detection}. Work of the first kind builds a bipartite graph based on URLs, hashtags, or other objects shared by accounts and estimates similarity between accounts \cite{pacheco2021uncovering}. A representative method of the second kind focuses on behavioral patterns. For example, Digital DNA \cite{cresci2016dna} encodes each account's activity as a string and compares strings across accounts. For the third, methods have been proposed that measure how far two accounts post in the same hours using dynamic time warping distance \cite{7837909}. There is also work that combines these elements \cite{keller2020political}.
    
    In practice, however, there is also behavior in which several accounts post the same wording, with at least one account posting that wording repeatedly, so that the same message is disseminated in a coordinated manner \cite{pacheco2020unveiling}. The behavior is not fully captured by this three-way scheme. First, it does not require shared external objects, such as common URLs or retweeted posts, because each account can independently post the same text. Second, it cannot be identified by activity type alone, as it may consist of human-like posting. Third, it does not require temporal synchronization, as identical wording may be repeated across multiple periods. Since what the accounts share is the text per se, we treat the repeated dissemination of identical wording by several accounts as coordination. We therefore treat textual identity as an additional shared trace alongside objects, activities, and timing.

    \subsection{Coordinated Activity in the Chinese-language space}
    In China, mass media and domestic social media are tightly controlled, and access to the outside internet is filtered by what is commonly called the Great Firewall \citep{Roberts+2018+190+222}. To reach information that does not circulate domestically, some Chinese speakers use foreign platforms \citep{Roberts+2018+190+222}. Prior reports have described parts of the Chinese-language space on X as containing substantial coordinated spam, including pornography and gambling solicitations \citep{dfrlab2024spambots}. Industry investigations have separately documented a political operation that camouflages state messaging inside spam-like account networks \citep{graphika2019spamouflage}. 
    
    This activity is not confined to the Chinese-language space. Reports indicate that some accounts become active in the Japanese-language stream around political events, even though most of their usual posts are in Chinese. These reports describe such accounts as mostly low quality \cite{shincho2026sns}. Because they write predominantly in Chinese, they are readily distinguishable from predominantly Japanese-language accounts. Consequently, these reports suggest that their direct influence on Japanese-language accounts has been considered limited. However, the reported cases focus on specific political events, whereas the overall volume of such activity is far larger. Whether this broader activity is primarily political, or whether it elicits reactions from users outside the coordinated population, remains unclear. We therefore study the Chinese-language space at the point where it meets a non-Chinese audience. Our target is not Chinese-language coordination on X in general, but rather Chinese-language accounts that surface in the Japanese stream on X.

    \subsection{Research Questions}
    
    The Chinese-language space is a particularly useful setting in which to examine how crude coordination operates and whether it reaches users outside its own population. Our study thus focuses on the following research questions:
    
    \begin{description}
    \item[RQ1] How can coordinated accounts, defined here as accounts whose output consists mostly of wordings that several accounts post and at least one of them repeats, be characterized? How long do they survive, and what are the temporal dynamics and semantic content of their posting behavior? 
    \item[RQ2] Are surges in coordinated activity organized around political events, as the flooding and information-operations literature would predict? What do coordinated accounts mainly post during surges?
    \item[RQ3] Within large observable cascades, do coordinated posts engage users outside their own population, and if so, which types of content generate this external interaction? 
    \end{description}

    \section{Methods}

    \subsection{Data Collection}
    To evaluate interaction between Chinese-language accounts and Japanese-language accounts, data from both user groups are required. We collected a full-volume archive of X posts spanning from May 1, 2024, to March 31, 2026 (JST), which X's proprietary, non-public classifier had labeled as Japanese. For posts receiving at least 1,000 aggregate reactions (including replies, quotes, mentions, and reposts), the archive captures all reactive posts alongside their referenced post IDs. Constructing directed links from each reactive post to its antecedent yields a cascade tree, which represents the full hierarchical reaction network originating from a single root post. Although root posts must receive at least 1,000 aggregate reactions to be included, reactions within their cascades are included regardless of how many reactions they themselves receive. Within this Japanese-classified stream, we identified numerous accounts that posted predominantly in Chinese. Using the procedure outlined below, we retrieved these accounts to analyze their behavioral characteristics and their interactions with Japanese accounts.
    
    \subsection{Selecting Accounts That Post Predominantly in Chinese}

    To identify Chinese-language accounts, we used the share of kana in each account's posts. Japanese typically uses both Han characters and kana together, whereas Chinese uses no kana. We therefore use a low kana share as a heuristic for identifying accounts that post predominantly in Chinese. For each account $i$, we concatenated all of its body posts within the window into a single string $T_i$ and computed Equation~\ref{eq:kana_ratio}, counting hiragana, katakana and halfwidth katakana as kana and the CJK unified ideographs as Han characters. We omitted Latin letters, digits, symbols and emoji. The distribution is markedly bimodal with a trough around 0.2, and moving the threshold anywhere between 0.2 and 0.4 changes the population by only about 2.3\%, so we set it at 0.2. Accounts whose posts contained neither kana nor Han characters, for which the ratio is undefined, were excluded and reported separately as leakage.

    \begin{equation}
    \mathrm{kana\_ratio}(i)
    = \frac{N_{\mathrm{kana}}(T_i)}
     {N_{\mathrm{kana}}(T_i) + N_{\mathrm{han}}(T_i)}
    \label{eq:kana_ratio}
    \end{equation}
    
    \subsection{Removing Accounts with Few Posts}
    
    Short Japanese texts can be written in Han characters alone, so an account with few posts could be misclassified on the kana ratio. When we drew only $k$ posts per account from accounts with a high kana ratio, the number misclassified at or below 0.2 decreased as $k$ increased and reached one at $k = 13$. We therefore require at least thirteen body posts. The population satisfying both conditions contains 1,065,150 accounts. We define these accounts as Chinese-language accounts.

    \subsection{Collecting Coordinated Texts}
    Since we define coordination as several accounts posting the same wording, with at least one account posting it repeatedly, we first normalized the texts posted by Chinese-language accounts and then identified the wordings that recur across and within accounts. Normalization is necessary because the same template often differs superficially between posts. Concretely, we removed emoji, URLs, mention handles, all whitespace including line breaks and tabs, and the invisible characters that arise from copying out of web pages; we lower-cased the alphabet; and we applied NFKC normalization, which unifies full-width and half-width forms. In addition, when the same character occurs three times or more in a row we shortened it to two. Through this normalization, the single most frequent wording came to account for 65,563,573 posts, of which only 2,327 (0.0036\%) matched exactly before normalization.
    
    \subsection{Identifying Coordinated Texts and Accounts}
    
    To define a coordinated text, we counted for every distinct wording the number of distinct accounts that posted it and the largest number of times any single account posted it. Most wordings were posted once by one account, while a small number of texts were posted by many accounts, each repeatedly. For both measures, the distribution shows a break at two. However, since two occurrences can arise by chance or by misoperation, we required both measures to exceed three. Texts meeting both conditions account for 96.02\% of the 515,793,498 body posts made by the Chinese-language population (495,279,985 posts); the two one-sided cases together account for 0.86\%.
        
    We then defined a coordinated account by how much of its output consists of coordinated texts. For each account, we took the share of its body posts that are coordinated texts, and treated the account as coordinated when that share was at least one half. The distribution of that share proved to be strongly bimodal, with a wide empty valley between the two modes. 63.4\% of accounts lie above 0.95 and 14.4\% below 0.05, while the entire range from 0.45 to 0.90 holds only 4.68\% of accounts, with no five-point bin holding more than 0.65\%. Any threshold placed inside that valley therefore separates the same two modes, and the choice of 0.5 is not consequential: moving it from 0.5 to 0.3 or 0.7 changes the coordinated-account share only from 68.9\% to 72.1\% or 66.8\%, respectively. This gave us 734,173 coordinated accounts. Applying the identical procedure to Japanese-language accounts over a single month gives 10.7\%, against 63.0\% for Chinese-language accounts under the same conditions (Appendix~\ref{app:baseline}).

    \subsection{Identifying Suspended Coordinated Accounts}
    To check our identification against the platform's own enforcement, we queried the X API for all 734,173 accounts we had identified as coordinated. Of these, 710,595 (96.79\%) had been suspended, while 22,394 (3.05\%) were still active (the remaining 0.16\% could not be retrieved). We also examined the suspension rate among non-coordinated Chinese-language accounts. We randomly sampled 57,200 accounts from 330,977 whole non-coordinated Chinese accounts. These results show that 12,584 (22.0\%) had been suspended, and 43,408 (75.9\%) accounts were active (the other 2.1\% not found). These results provide external corroboration rather than ground truth. Suspension reflects X's enforcement decisions under its own rules; the reasons for individual suspensions are not public, so these figures should not be interpreted as measures of classification accuracy.

    \subsection{Identifying amplifier and reactor accounts}
    
    Related to coordination, we further define two actors: amplifier and reactor. Amplifier accounts repost the posts of coordinated accounts and reactor accounts reply to, mention or quote them. In both cases, we require that these actions occur more often than expected by chance. We identify them using exact binomial tests against non-coordinated baseline reaction rates, with Benjamini--Hochberg false discovery rate corrections across accounts. We additionally require that a majority of the account's actions of that type be directed at coordinated accounts. The base rates, the test and its sensitivity to both thresholds are given in  Appendix~\ref{app:amplify_reactor}.

    \subsection{Topic modeling and annotation with Language Models}
    
    \subsubsection{Embedding}

    We analyzed the distinct normalized texts of the whole Chinese-language population, not only of coordinated accounts, so that what the two groups post can be compared within one set of categories. Working on distinct texts rather than posts keeps the space from being dominated by whichever template was posted most often; post volume re-enters only when results are weighted for reporting. Texts were embedded with multilingual-e5-base, which produces 768-dimensional L2-normalized vectors \citep{wang2024multilingual}. We used a multilingual encoder because it supports both simplified and traditional Chinese, which are both present in the corpus. We then reduced the embeddings to five dimensions with UMAP \citep{healy2024uniform} under the cosine metric, because density-based clustering degrades in high-dimensional spaces where distances concentrate; five dimensions is the conventional setting in BERTopic-style pipelines \citep{grootendorst2022bertopic}.

    \subsubsection{Topic Modeling} 
    
    The pipeline has three stages. Clustering surveys what kinds of text are present; the model then writes a category scheme from what the clusters contain; and the scheme, once frozen into a codebook, is applied to individual wordings to estimate composition. Only the third stage produces a reported number. First, the reduced vectors were clustered with HDBSCAN \citep{campello2013hdbscan}. The minimum cluster size was set to 100 texts; all other parameters were left at their default values. The procedure yielded 819 topics, of which 750 contain at least one post written by a coordinated account; only these were carried forward. The result has two properties that constrain how it may be used. The cluster structure was learned from a uniform sample of 1,200,000 of the 14,481,385 texts, within which HDBSCAN marked 57.7\% as noise. Labels were then propagated to every text by nearest centroid without a similarity threshold, which absorbed the noise points as well, so a topic's periphery is attached by proximity alone and its label describes only its core. We therefore use the topics only to build the codebook and to survey what kinds of text exist, and we estimate composition at the level of the text (Appendix~\ref{app:codebook}).

    \subsubsection{Annotation with Language Models} 
    
    The large language model (LLM) is used in two distinct ways. Both uses were run with Claude Haiku 4.5. It first induces a set of categories, working from summaries of the 750 topics rather than from the texts themselves: it writes a label and a definition for each topic, and those 750 labels are then merged into a taxonomy. No category set exists before this step. Once that taxonomy has been turned into a codebook, the LLM assigns units to it zero-shot. The unit at this stage is the individual wording, not the topic, and it is these assignments that produce the composition we report (Appendix~\ref{app:codebook}). The prompt carries the category names, decision rules and checking order, but no labelled examples. Zero-shot assignment is an established strategy in the social sciences \citep{gilardi2023chatgpt, ziems2024can}, on the condition that the labels are validated rather than assumed correct \citep{pangakis2023automated}. This setting makes validation particularly important: the coding scheme was induced from the same corpus by the same model, so an error in the categories cannot be detected by the assignment step that uses them. We therefore assess both the reliability of the LLM assignments and their agreement with human coding, as described below.

    \subsubsection{Labelling and Category Induction} 
    
    First, each of the 750 topics was summarized for the prompt by its ten most distinctive terms (c-TF-IDF), its fifty most frequent texts with their post counts, and ten texts drawn at random; the frequent texts show what the volume consists of, while the random draw reveals content beyond the most frequent texts. Following \citet{pham-etal-2024-topicgpt}, the model returned a label of at most three words, a one-sentence definition, and a verbatim quotation from the supplied texts as evidence, and was instructed to introduce no concept the texts do not support. 
    
    Second, the 750 labels and definitions were then merged into higher-level categories following the generate-then-refine procedure of \citep{pham-etal-2024-topicgpt}. Ten batches ordered by post volume were prompted independently for candidate categories defined by what a post functionally does, yielding 66 candidates, which a single refine prompt merged into a taxonomy of at most ten mutually exclusive categories with exactly one residual. The result was a draft of nine categories (Table~\ref{tab:draft}).
    
    \subsubsection{Codebook Construction} 
    
    Following standard practice in content analysis \citep{krippendorff1989content}, we froze the refined categories into a codebook before coding began. Each category is defined by five fields: a name, a one- or two-sentence definition, an Include list of textual signals that place a text in the category, an Exclude field that names the nearest competing category and states which of the two a borderline text belongs to, and a decision rule that can be applied to a text on its own (Appendix~\ref{app:frozen-codebook}). Definitions that appealed to what an author intended were rewritten as criteria observable in the text per se, namely whether the text solicits or directs the reader, what it refers to, and whether it is a repeated template.

    \subsubsection{Resolving Texts That Satisfy More Than One Rule} 
    A single text can satisfy multiple decision rules. For example, a post promising trading profits while blaming a named government meets both the money-solicitation and political-claims criteria. To address such overlaps, the codebook establishes a fixed priority order, assigning a post to the first category whose rule it satisfies. The evaluation order proceeds from the rarest to the most generic categories: illegal material, insults and abuse, political claims, money solicitation, sexual and dating solicitation, requests to like or follow, praise and good wishes, No recoverable message, and other everyday topics. Because political claims are evaluated before money solicitation, the low political share we report is not an artifact of this classification scheme (Appendix~\ref{app:checking-order}).

    \subsubsection{Codebook Refinement} 
    
    The codebook went through two revisions before being frozen. The first draft mixed two questions in one variable---whether a text is a mass-produced template, and what it is about---and a post can be both; separating them, replacing the template category with No recoverable message, and removing a ``mixed'' option that let the model escape the decision, raised inter-run agreement at the topic level from $\kappa = 0.528$ to $\kappa = 0.660$ (Appendix~\ref{app:two_revision}).

    \subsubsection{Units of Analysis} 

    The frozen codebook is then applied at two levels. At the topic level we classified each of the 750 topics; this is the level at which the revisions above were diagnosed and at which the $\kappa$ we report was measured, and it serves only to characterize what kinds of topic exist. The composition we report comes from the second level, at which we classified distinct wordings directly. The 495,170,607 body posts of coordinated accounts reduce to 545,291 distinct wordings, of which the 5,000 largest by post volume carry 98.33\% of the posts; we classified all 5,000 and estimated the remaining 1.67\% from a uniform random sample of 1,000 tail wordings. Each wording was classified in three independent runs and assigned by majority; where all three disagreed it was recorded as ``undecided'' and its volume reported under that label. Pairwise agreement across runs was 76.0\% for the 5,000, with three-run unanimity on 65.6\%. Every composition figure in the Results is a text-level estimate weighted by post volume.
    
    \subsubsection{Reliability Assessment} 
    
    With the codebook frozen, every unit was classified independently in separate single-turn runs with no shared state. The display order of the categories in the prompt was shuffled per batch from a fixed seed, since LLMs are sensitive to option order \citep{pezeshkpour-hruschka-2024-large}. This is independent of the fixed classification hierarchy, which is stated in the decision rules. Agreement is reported as Cohen's $\kappa$ \citep{cohen1960coefficient}.

    \subsubsection{Human Validation} 
            
    Two of the research group members independently coded a stratified subsample using the frozen codebook. Neither was informed of the purpose of this research. The subsample consisted of 26 wordings drawn from the head of the post-volume distribution, 220 drawn equally across the nine categories and the undecided label, and 4 drawn from wordings containing a term from a fixed political vocabulary list, since a systematic failure to recognize political content would depress the political share without depressing agreement on the categories that fill the corpus. Inter-coder agreement was $\kappa = \mathbf{0.543}$, and agreement between the agreed human label and the LLM label was $\kappa = \mathbf{0.421}$. Classification agreement varies substantially across categories. Among the items on which the two coders agreed, every post the LLM labelled sexual solicitation was confirmed (16/16), but only 31.1\% (14/45) of those it labelled illegal material under the codebook rule. Of the 45 posts labelled illegal material, the coders assigned 31 to sexual solicitation, of which only four carry terms denoting minors; the rest are adult pornography the LLM over-labels. The 14 left in the category are dominated by the illicit trade of personal data. Replacing the LLM labels with the human labels would move the political share from 0.038\% to 0.035\%, leaving the rank ordering of categories unchanged.

    \section{Results}

    \subsection{Characteristics of Coordinated Accounts}

    \begin{figure*}[!t]
    \centering
    \includegraphics[width=0.95\textwidth]{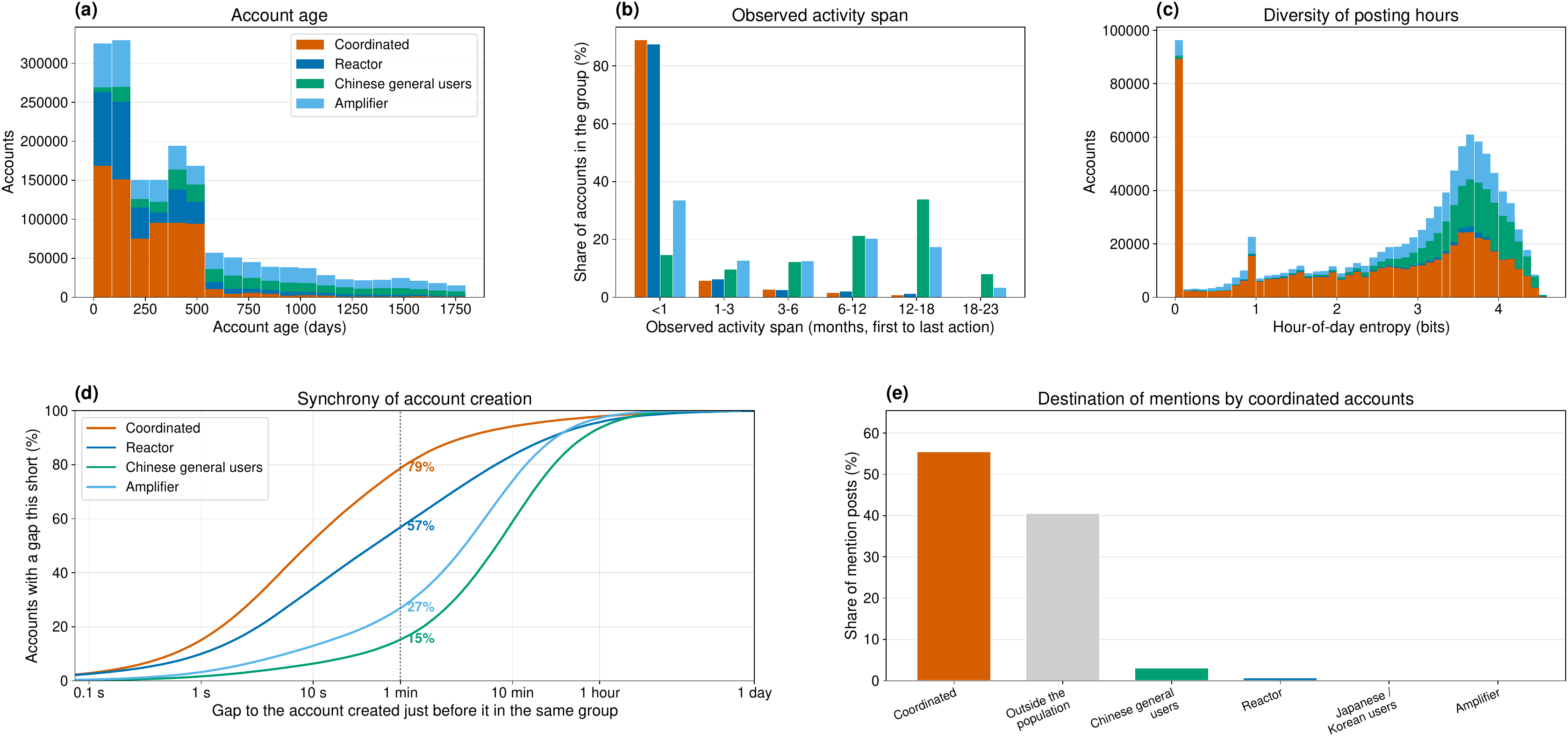}
    \caption{Behavior characteristics}
    \label{fig:behavior-char}
    \end{figure*}
    
    We first document the behavior patterns of coordinated accounts to answer RQ1. Figure~\ref{fig:behavior-char} shows five characteristics of these accounts: account age, activity span, the temporal concentration of posting, the timing of account creation, and the destinations of their mentions. Coordinated accounts are young and short-lived. Their median account age, recovered from the Snowflake identifier and measured to the end of the study period, is 237 days, against 824 days for Chinese general users. More strikingly, the median interval between an account's first and its last post -- what we call its lifespan -- is 0.06 months, roughly 1.8 days, against 10.18 months for Chinese general users. These accounts are not merely new; they are disposable. 
    
    To measure how unevenly their activity is spread over the day, we compute the information entropy of each account's posting hours: posts are binned into the 24 hours of the day in JST and the base-2 entropy of that distribution is taken, for accounts with at least 24 posts. Coordinated accounts are the most concentrated of the four groups, at a median of 2.69 bits against 3.69 for Chinese general users, 3.41 for amplifiers and 3.21 for reactors. The separation is real but narrow -- under one bit -- so this measure alone does not distinguish the groups. We also demonstrate how these coordinated accounts were created within a similar timeframe using Snowflake IDs. The results indicate bulk registrations occurring within extremely short intervals, with a median gap of only 8.9 seconds between account creations. 
    
    Finally, we document the target diversity of mentions: 69.20\% of all their text-bearing actions -- original posts, replies and mentions, quotes posts taken together -- are mention posts. Since an account that mentions more will inevitably repeat recipients, we compared accounts of similar volume: among those that sent at least a thousand mentions, the median coordinated account reached just 9 distinct recipients out of 2,465 mention posts, while the median Chinese general users reached 428 out of 1,550. Coordinated accounts thus mention an extremely narrow set of recipients, over and over: 55.44\% of their mentions are addressed to other coordinated accounts, against 4.73\% for Chinese general users, who send 79.58\% of their mentions to other general users. More than half of their mentions thus circulate within the coordinated group. The further 40.55\% falls outside the Chinese-dominant population and represents a channel through which this activity reaches users outside that population, an asymmetry we return to in RQ3. The 30 most frequent texts account for 40\% of all coordinated posts, and all are mention posts addressed to specific users, consistent with the closed mentioning pattern described above. The top three are variants of one text and alone carry 141 million posts (28.5\%; Table~\ref{tab:top-coord}).
    
\begin{figure*}[!t]
    \centering
    \includegraphics[width=0.95\textwidth]{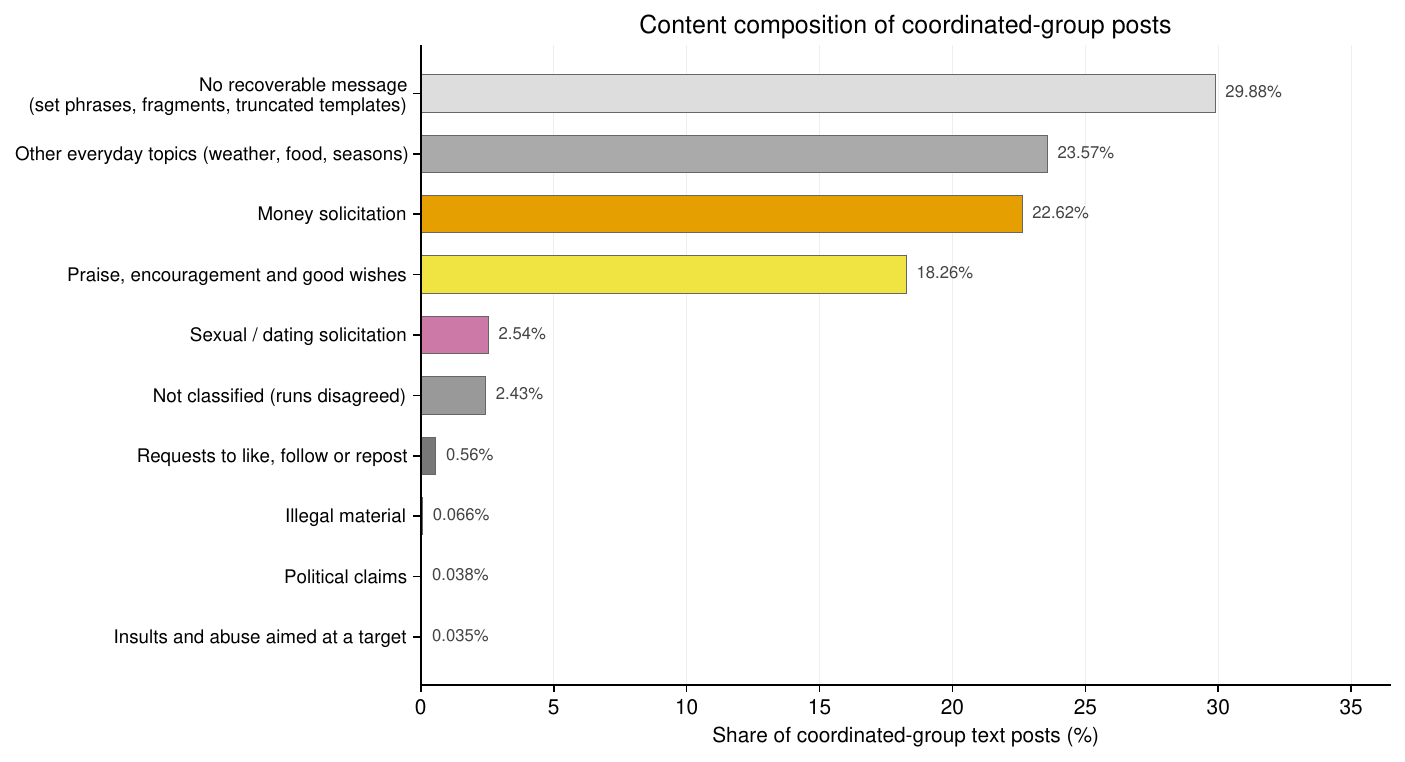}
    \caption{Content composition of coordinated-group posts}
    \label{fig:composition}
    \end{figure*}

    Next, we present what kinds of posts appear in the output of coordinated accounts. Figure~\ref{fig:composition} gives the content composition of all 495,170,607 body posts made by these accounts, estimated at the level of distinct wordings and weighted by post volume. Note that the unit here is the account's entire output and not only its coordinated texts. The codebook has nine categories, and what fills the corpus is concentrated in four of them. The largest is No recoverable message---set phrases and fragments that carry no message---at 29.88\%, followed by other everyday topics such as weather, food and the seasons at 23.57\%, money solicitation (get-rich-quick offers, investment and part-time-work pitches, and outright fraud) at 22.62\%, and praise and good wishes at 18.26\%. Sexual and dating solicitation accounts for 2.54\%, and political claims for 0.038\%, fewer than four posts in every ten thousand. Most of what coordinated accounts post is therefore not political: what fills their volume is money solicitation and the repetition of content-free strings.
    
    \subsection{Surges in Coordinated Activity}
    \begin{figure*}[!t]
    \centering
    \includegraphics[width=0.75\textwidth]{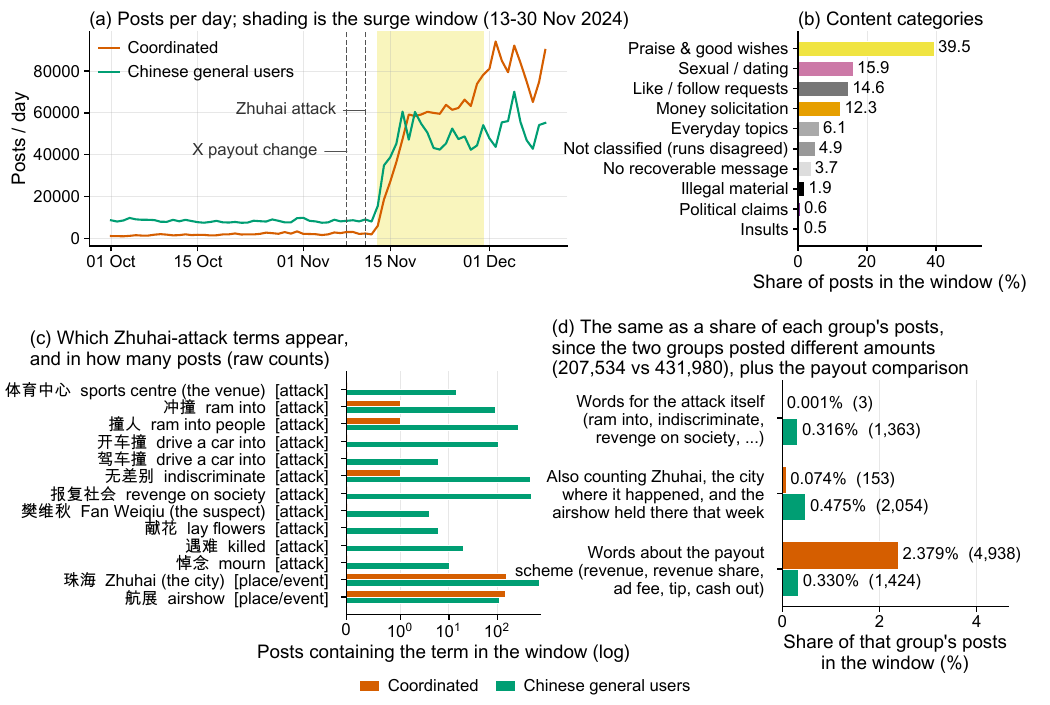}
    \caption{The November 2024 surge in coordinated posting}
    \label{fig:surge-anatomy}
    \end{figure*}

    \begin{figure*}[!t]
    \centering
    \includegraphics[width=1.0\textwidth]{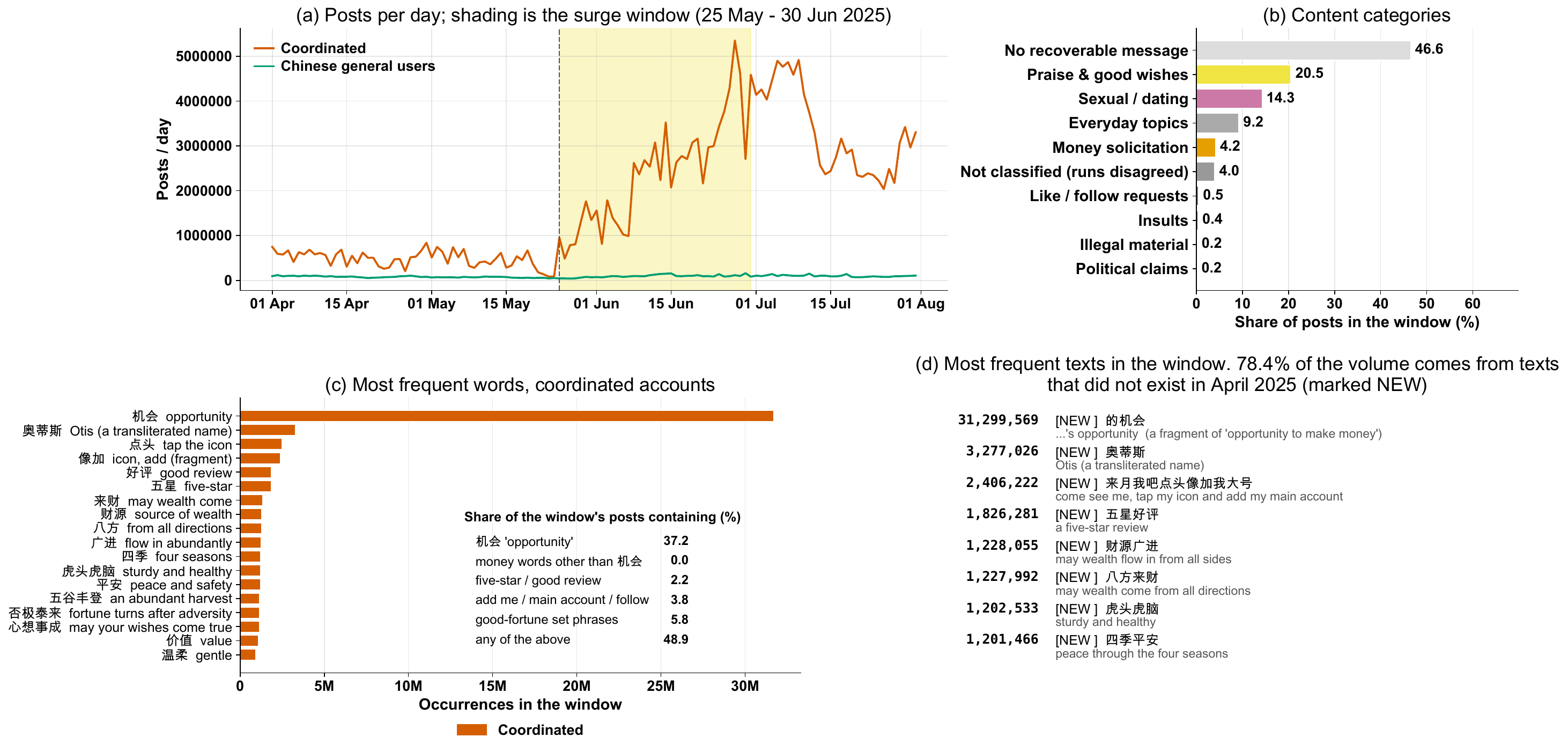}
    \caption{The May–July 2025 surge in coordinated posting}
    \label{fig:surge2025}
    \end{figure*}
    
    Next, we analyze whether two substantial surges were organized around political events to answer RQ2. Figures~\ref{fig:surge-anatomy}(a) and \ref{fig:surge2025}(a) show daily posting volume around the two observed surges in November 2024 and May--July 2025. During the first surge, both coordinated accounts and general Chinese accounts show a simultaneous increase in activity. In addition, we identified two candidate cases in November 2024: one political event and one change to X's policy. First, on 11 November 2024, 35 people were killed and 48 others injured by a car in a random attack in Zhuhai, China \cite{gan2024zhuhai}. This was the deadliest attack in China since May 2024 \cite{gan2024zhuhai}, and it is said that after this case information control became severe in order to prevent copycat crimes \cite{apnews2025deadattack}. Second, we found that a change to X's payout policy had been announced in the same period \cite{x_creator_revenue_sharing_2024}. This update marks a fundamental shift from a payout model based on ad impressions in the reply section to one driven by engagement -- likes, replies, reposts and so on -- from X Premium users. The change takes effect on 8 November, followed by the attack on the 11th and the surge on the 13th. We applied the same text analysis as above to the surge window. Figure~\ref{fig:surge-anatomy}(b) shows what these accounts were posting during the 2024 surge. Praise and good wishes take 39.5\% of the window, sexual or dating solicitation 15.9\%, requests to like or follow 14.6\%, and money solicitation 12.3\%, while political claims take 0.60\% and insults 0.50\%. In other words, what fills the window is the material that generates engagement and asks for money, not political argument. Content classification cannot, however, detect flooding, in which posts carry the vocabulary of an event without arguing about it and so crowd its search results \citep{Roberts+2018+190+222}. We therefore also counted words, using two lists fixed before counting. The narrow list holds only words for the attack itself -- the venue, ram into, drive a car into, indiscriminate, revenge on society, the suspect's name, killed, mourn, lay flowers. The broad list adds the name of the city, Zhuhai, and the name of the airshow held there in the same week, because a person searching for the attack would meet both, and a post carrying either occupies the same result page. These results indicate that coordinated accounts did not dominate the discourse surrounding the attack (Figure~\ref{fig:surge-anatomy}(c)). Because coordinated activity consists mostly of reposts, only 207,534 of its 962,419 actions in the window carried text, against 431,980 of 829,370 for Chinese general users; we use these text posts as denominators. Only 3 of the coordinated text posts (0.001\%) contained target attack keywords, compared to 1,363 (0.316\%) for Chinese general users. Even on the broad list, they reach only 0.074\% against 0.475\%. Words about the payout scheme move the other way: 2.379\% of their posts against 0.330\%, a factor of seven (Figure~\ref{fig:surge-anatomy}(d)). These results do not show the cause of the discontinuity, but they constrain its content: the surge shows little evidence of flooding around the attack or of spreading political claims, while its vocabulary is more consistent with the payout change.
 
    During the second, coordinated accounts surged while Chinese general users remained at baseline levels. In this period, we found no corresponding event in Chinese or Japanese society and no relevant change in X's policies that we could find. Figure~\ref{fig:surge2025}(a) shows that it begins on a single day: daily posting by coordinated accounts rises from 85 thousand on 24 May to 953 thousand on 25 May and keeps increasing until it peaks in July, so we take 25 May to 30 June as the surge window. The surge is not an artefact of collection, since monthly posting by accounts outside the Chinese-dominant population moves only within 20\% while the coordinated share of all posts rises from 5.56\% in April to 22.75\% in June; amplifiers follow weakly and with delay (2.16 times baseline against 7.42), and reactors not at all (0.92).

    Figure~\ref{fig:surge2025}(b) shows No recoverable message at 46.6\% of the window, praise and good wishes at 20.5\%, sexual or dating solicitation at 14.3\%, and political claims at 0.2\%. This understates money solicitation: the most frequent text, with 31.3 million posts, is a truncated form of ``an opportunity to make money'' that reads only as ``an opportunity of ...'' and is therefore classed as No recoverable message, and texts containing ``opportunity'' account for 37.2\% of the window against the 4.2\% the categories report, while money vocabulary other than that token covers 0.0\% (Figure~\ref{fig:surge2025} (c)). Texts absent in April account for 78.4\% of the volume, but this is turnover within one family rather than a new operation: texts containing ``opportunity'' rise from 0.0\% of coordinated posts in March to 37.6\% in June while the individual strings are replaced month by month, as would be consistent with an operation in which individual strings are regularly replaced. We can therefore say what this surge consisted of---the money-solicitation family, and not political claims---but not what drove it, and we keep it separate from the November 2024 discontinuity, where timing and content both coincide with the payout change.

    \subsection{Cascade Analysis}
    
    \begin{figure*}[!t]
    \centering
    \includegraphics[width=0.90\textwidth]{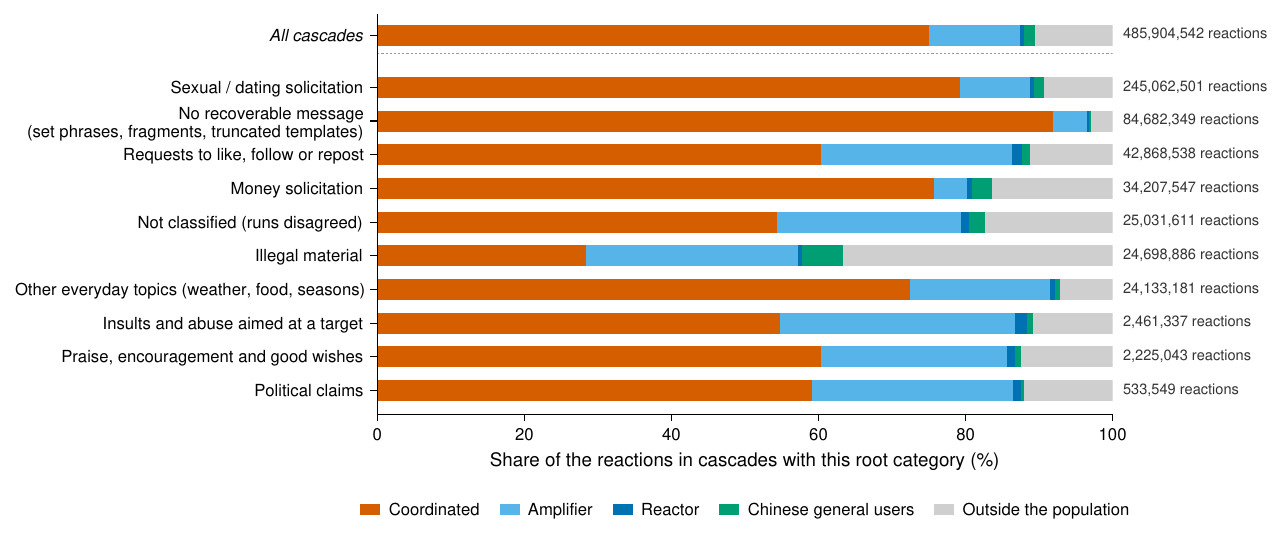}
    \caption{Who reacts, by the content of the root post}
    \label{fig:cascade-reach}
    \end{figure*}

    Finally, we present how coordinated posts elicit interactions from non-coordinated accounts to answer RQ3. To answer this, we examine two key aspects of the cascade trees: (i) how deeply coordinated posts elicit reactions from accounts outside the coordinated population, and (ii) which types of posts elicit reactions and who those respondents are. Our cascade data are built from the reaction trees of root posts that drew at least a thousand reactions. A tree deepens only when a user writes in their own words, that is, replies or quotes, because a repost adds one level and cannot itself be replied to or reposted. There are 116,532 such trees rooted in coordinated posts, and they contain 485,904,542 reactions in total (Figure~\ref{fig:cascade-reach} top bar). Of those reactions, 290,861,721 (59.9\%) lie at depth one and 194,400,568 (40.0\%) at depth two, so 99.9\% stay within two levels of the root  (Figure~\ref{fig:cascade-dist} (e)). The two levels are not the same act: mentions and replies account for 86.7\% of the reactions at depth one but only 4.5\% at depth two, where reposts are 95.5\%. Over the whole tree, 53.7\% of the reactions are mentions or replies and 46.2\% reposts. Most cascades therefore remain shallow. Compared with cascades whose root was written by a Chinese general user, coordinated cascades are markedly shallow---mean maximum depth 1.494 against 4.034, 64.5\% ending at depth one against 32.1\%, and 0.75\% reaching depth five or more against 19.5\% (Figure~\ref{fig:cascade-dist} (a--b)). Given the shallow cascade depth and the high tendency toward self-referential mentions among coordinated accounts, within these large observable cascades, coordinated posts primarily circulate within the coordinated population rather than reaching accounts outside it.

    Next, we present which types of coordinated posts draw more reactions and who they reach. To analyze these dynamics, we applied the same frozen codebook to the root post of all 116,532 coordinated root cascades. Each of the 30,396 unique root wordings was assigned to one of nine content categories (or an ``undecided'' option) over two independent classification runs using Claude Haiku 4.5. Any inter-run disagreements were adjudicated by a third run using majority rule. The two runs agreed on 65.3\% of wordings (Cohen's $\kappa = 0.487$), and 5.4\% of cascades remained undecided. Examining cascade depth shows that, regardless of category, most cascades reach a depth of less than three (Figure~\ref{fig:cascade-topic}). Coordinated posts are mainly reacted to by coordinated accounts; however, posts in the illegal materials category are relatively frequently reacted to by non-coordinated accounts. In Figure~\ref{fig:cascade-reach}, the categories are ordered in descending order based on the number of reactions each category received. The results show that sexual/dating solicitation garners the most reactions, followed by No recoverable message, request to like, follow or repost, and everyday topics. Regarding who reacted, posts in the three categories above primarily circulate within coordinated accounts, as is the case for most other categories. In contrast, illegal materials notably garner reactions from non-coordinated accounts. In particular, 40.7\% of total reactions came from outside the Chinese-speaking population, representing a comparatively high proportion of external reactions across categories. Because this group is defined by exclusion, it represents an upper bound on engagement by non-coordinated accounts. Since the category boundary itself proved imprecise, we asked whether this reach follows from the illegality of the content or from its explicit vocabulary. Within the illegal-material category, roots carrying terms that denote minors reach outside at 44.1\% and roots carrying adult sexual terms at 41.2\%, a difference too small to attribute the reach to illegality. The same vocabulary raises reach inside the sexual solicitation category, from 10.7\% for that category as a whole to 36.2\% for the roots that carry adult sexual terms. These results suggest that external reach is more closely associated with explicit vocabulary than with the assigned category label (Appendix~\ref{app:vocab-reach}). The volume of reactions this category receives is also substantial (24,698,886).

    \section{Discussion}
    
    Although Generative AI is expected to make coordinated accounts cheaper, more personalized, and more convincing \citep{goldstein2023generative, ferrara2024genai, sison2024chatgpt}, the behavior we document reflects crude automation. The crude coordinated accounts observed here, some of which may be bots, form a largely self-referential ecosystem within social media: although a substantial share of their mentions is addressed outside the network, the reactions they receive come mostly from within it.

    This interpretation is further supported by our analysis of the surges in coordinated activity, where amplifiers followed the coordinated accounts only weakly and with delay and reactors did not follow at all. Within the scope of our dataset, crude coordination is predominantly commercial, with political claims accounting for merely 0.038\% of all coordinated posts (Figure~\ref{fig:composition}). Regarding the two major surges in coordinated activity, we found no clear relationship with external political events, and only the first coincided with a change to X's payout rules. Instead, this activity is consistent with financial incentives, given its temporal coincidence with changes in the platform's monetization policy and the characteristics of the posts. One plausible interpretation is that this coordination was intended, at least in part, to inflate engagement metrics through intra-network mentions, although our analysis does not establish that monetization policies caused the activity.

    The cascade analysis shows that although most coordinated posts fail to elicit discussion among general users, one kind of content reaches outside the Chinese-speaking population at the several times the rate of the rest. That kind is defined not by the category boundary but by the vocabulary: posts carrying explicit sexual or illicit-trade terms reach outside at around 40\% whether they fall under illegal material or under sexual solicitation. This reading does not depend on the category label, which our human validation showed to be imprecise. Regarding suspension rates, 710,595 out of the 734,173 coordinated accounts (96.79\%) had been suspended by X at the time of our query. However, these suspension rates vary across categories. Accounts whose primary content consisted of political claims were suspended at a rate of 100\%, whereas those whose primary content was classified as illegal material had a lower observed suspension rate of 91.2\% (Figure~\ref{fig:suspension}(a)). Suspended accounts were created more recently, had shorter lifespans, and posted on more rigid schedules than those not suspended (Figure~\ref{fig:suspension}(b--d)).

    Taken together, the cascade findings and suspension rates suggest a way of prioritizing detection. Although political content has long been a primary focus of detection research, the posts that reach furthest outside the coordinated population are those carrying explicit sexual or illicit-trade vocabulary, and the category holding most of them is suspended less often than most others. Detection keyed to that vocabulary rather than to a category label may therefore be relevant to future research on platform moderation.

    Our study has five limitations. First, our target population is constrained by two methodological exclusions. Our dataset comprises posts classified as Japanese by X. Because the classifier is not part of our study design, we cannot independently characterize its inclusion and exclusion criteria. If the classifier disproportionately categorizes short, Han-character-only, or mention-heavy texts as Japanese, our dataset overrepresents these formats. Consequently, the observed low proportion of political content and short account lifespans may reflect coordination patterns unique to this specific subset. Within this stream, we then isolate accounts posting predominantly in Chinese. Accounts that post in Japanese only occasionally are retained, since a few Japanese posts barely move the kana ratio. However, operations predominantly using the Japanese language failed to meet our inclusion threshold, placing previously reported influence campaigns on Japanese domestic politics outside the scope of this study. Extending this methodology to Japanese-language output remains a topic for future research. 
    Second, our analysis is limited to a single political event and one monetization update; moreover, because Chinese authorities reportedly suppressed news coverage of similar attacks, public discussion surrounding these events may be artificially subdued in our dataset \cite{apnews2024zhuhai, apnews2025deadattack, cnn2025zhejiang}, so unobserved political events cannot be ruled out. 
    Third, we cannot rule out a data-collection artifact behind the November 2024 surge: Chinese general users also rose roughly sevenfold that day, against a 1.0- to 1.2-fold change in non-Chinese accounts, and the content differences between the two groups do not exclude a classification shift that would carry both. The November 2024 findings therefore remain tentative. 
    Fourth, our content categories rely on language-model annotations. Human validation covered only on a stratified subsample, so agreement beyond it is assumed. Agreement was also uneven across categories: of the 45 wordings the LLM assigned to illegal material, the coders reassigned 31 to sexual solicitation, so the illegal-material share we report is likely inflated. This is why the reach result above is stated in terms of vocabulary rather than of the category label. Cases where three classification runs disagreed were labeled as ``undecided''. Conversely, the group outside our target population is defined by exclusion, and therefore contains not only Japanese-language users but also Chinese-language accounts below our posting threshold, representing an upper bound on ordinary-user engagement. 
    Fifth, our detection method relies on text normalization to identify identical posts. Consequently, sophisticated coordination that varies its phrasing--such as content generated by AI--remains undetected, limiting our scope to relatively crude coordination. Note, therefore, that our reported 68.9\% represents the share of Chinese-language accounts engaged in such crude coordination, rather than the proportion of all coordination that is crude. The same criterion applied to Japanese-language accounts over a single month yields 10.7\%, so the criterion is not one that any large population satisfies.

\section{Acknowledgments}
This paper is based on results obtained from a project, JPNP22007, commissioned by the New Energy and Industrial Technology Development Organization (NEDO).

$\,$

\bibliographystyle{plainnat}
\bibliography{aaai2026}



\newpage

\appendix
\setcounter{secnumdepth}{2}
\renewcommand{\thefigure}{A\arabic{figure}}
\renewcommand{\thetable}{A\arabic{table}}
\setcounter{figure}{0}
\setcounter{table}{0}

\section{Data and Coordination Detection}

\subsection{Amplifier and reactor accounts}
\label{app:amplify_reactor}
Aggregating over all accounts outside the coordinated accounts, 1.83\% of reposts, 0.50\% of replies and mentions, and 0.005\% of quotes were directed at coordinated accounts. We define these proportions as the base rates, denoted by $p_0(t)$, at which an action of type $t \in \{\mathrm{repost}, \mathrm{reply/mention}, \mathrm{quote}\}$ targets a coordinated account by chance. For each account $i$, let $n_i$ represent its total number of interactions of type $t$, and $k_i$ the subset directed at coordinated accounts. Under the null hypothesis $X \sim \mathrm{Binom}(n_i, p_0(t))$, we compute the upper-tail probability of observing $k_i$ or more such interactions, adjusting for multiple testing across accounts using the Benjamini--Hochberg procedure ($\alpha = 0.05$). However, statistical significance alone is insufficient; accounts with a large $n_i$ may achieve significance even with a negligible bias, which does not necessarily reflect deliberate targeting. Consequently, we additionally require that a majority of an account's actions of type $t$ be directed at the coordinated population---that is, $s_i = k_i / n_i \ge 0.5$.

\section{Topics, Codebook and Annotation}
\label{app:codebook}

\subsection{What the clusters can and cannot support}
Pairwise operations over the 14,481,385 texts of the clustering corpus are infeasible, so HDBSCAN was run on a uniform sample of 1,200,000 texts; that size was set by memory, being roughly the largest sample our hardware could hold. Within the sample, HDBSCAN marked 692,334 texts (57.7\%) as noise. The labels were then propagated to the rest of the corpus by attaching every text to its nearest cluster centroid, with no similarity threshold. This does not resolve the noise: a noise point is not given a group of its own but is forced into whichever cluster lies nearest, however far that is, and the resulting mapping contains no noise label at all.

A topic therefore has two parts. Its dense core was found in the sample; its periphery was attached by proximity alone. The label describes the core and makes no promise about the periphery.

This construction has a cost, and we measured it. Of the post volume in topics labelled sexual solicitation, only 10.4\% is sexual solicitation when the texts are classified one at a time; everyday topics account for 35.5\% and no recoverable message for 34.9\%. This is the gap between the topic-level figure of 18.64\% for sexual solicitation and the text-level figure of 2.54\% reported in the main text.

The cost falls on counting, not on the codebook. Each label was written from a topic's fifty most frequent texts and its most distinctive terms, that is from the core and not the periphery, and the nine categories were induced from those 750 labels alone. Absorption changes which topic a text is counted under; it does not change whether a kind of text was found, named, and carried into the codebook. no recoverable message is a case in point: it sits in the periphery of the sexual solicitation topics, and it also holds fifty topics of its own, under which it was named.

\subsection{The frozen codebook}
\label{app:frozen-codebook}
The frozen codebook—comprising five fields across nine categories and the explicit classification hierarchy—is available in an anonymized repository: \url{https://drive.google.com/drive/folders/1l_dt7c6HbQIeBofLzAl2dpmgwu7LqRq-?usp=sharing}. Table~\ref{tab:codebook-summary} gives the categories and their decision rules in brief.

\begin{table}[htbp]
\centering
\small
\begin{tabular}{p{0.28\columnwidth}p{0.62\columnwidth}}
\toprule
Category & Decision rule \\
\midrule
Illegal material & A reference to a minor, to non-consent, to illegal drugs or to weapons co-occurs with wording that offers or seeks access. \\
Insults and attacks & The target is identifiable from the text and words of contempt or threat are present. \\
Political claims & A political actor or a political issue is referred to. \\
Money solicitation & Wording about money, earnings or remuneration is present. \\
Sexual or dating solicitation & Wording about dating, sexual contact or courtship is present. \\
Requests to like or follow & The text asks the reader to act, with neither a money nor a dating context. \\
Praise and good wishes & Intelligible encouragement, praise or blessing, with no request or solicitation. \\
no recoverable message & No message can be recovered from the text. \\
Other & A message can be recovered but none of the eight preceding features is present. \\
\bottomrule
\end{tabular}
\caption{The nine categories and their decision rules, in the order in which they are checked.}
\label{tab:codebook-summary}
\end{table}

\subsection{A fixed classification hierarchy eliminates this ambiguity}
\label{app:checking-order}

Requiring categories to be mutually exclusive does not automatically make them so. Without a clear resolution rule, a text satisfying multiple criteria causes artificial disagreement, as different coders may validly assign different labels to the same post. A fixed classification hierarchy eliminates this ambiguity. The sequence remains identical across all prompts and runs, operating not as a measure of best fit, but as a systematic protocol. By prioritizing rare categories over generic ones, this order prevents rare themes from being absorbed into broader classifications.

\subsection{Codebook development and refinement}
\label{app:two_revision}

\begin{table*}[t]
\centering
\small
\begin{tabular}{p{0.24\textwidth}p{0.26\textwidth}p{0.40\textwidth}}
\toprule
Draft category & Frozen category & Change \\
\midrule
Illegal material & Illegal material & Kept; definition rewritten \\
Harassment & Insults and attacks & Renamed and redefined \\
Political propaganda & Political claims & Renamed and redefined \\
Money fraud & Money solicitation & Renamed and redefined \\
Sexual exploitation & Sexual or dating solicitation & Renamed and redefined \\
Engagement inducement & Requests to like or follow & Renamed and redefined \\
Trust building & Praise and good wishes & Renamed and redefined \\
Low-quality spam & --- & Removed; became the binary repetition attribute \\
--- & no recoverable message & Added \\
Other & Other & Kept; definition rewritten \\
\bottomrule
\end{tabular}
\caption{The nine draft categories and the nine frozen categories. Six were
renamed and redefined, two were kept, one was removed, and one was added.
Every definition that appealed to an author's intent was rewritten as a
criterion observable in the text.}
\label{tab:draft}
\end{table*}

Applying the nine-category draft to all 750 topics in two independent runs gave $\kappa = 0.528$ with 60.1\% raw agreement, and 36.1\% of post volume received different categories in the two runs. Every large disagreement involved one category, low-quality spam. The reason is structural. That category asks whether a text is a mass-produced template, while the remaining categories ask what a text is about. A single set of mutually exclusive categories cannot answer two questions at once, because a post can be both a template and a solicitation, and which label a coder returns then depends on which question they happened to weigh.

We therefore separated the two questions into two variables. Content is the eight remaining categories plus No recoverable message, assigned by the model. Repetition is a binary attribute computed from the post counts already supplied in the prompt: a topic is marked as templated when its single most frequent text accounts for at least 50\% of its posts, or its eight most frequent texts account for at least 80\%. Removing the form category from the content variable and computing it mechanically raised agreement to $\kappa = 0.629$ (67.7\% raw).

A second defect surfaced afterwards. The assignment prompt had been offering ``mixed'' as an option, which is not a category in the frozen codebook, so a coder could escape the decision the codebook required. Removing it and re-running all 750 topics raised agreement to $\kappa = 0.660$ (70.8\% raw), with the share of post volume receiving different categories falling from 19.59\% to 15.11\%. The codebook was then written to a file with the status FROZEN and the explicit rule that its wording is not edited once assignment begins.

\section{Additional Results}
\label{app:results}

\subsection{Cascade depth by root group}
\label{app:cascade-dist}
\begin{figure}[htbp]
\centering
\includegraphics[width=\columnwidth]{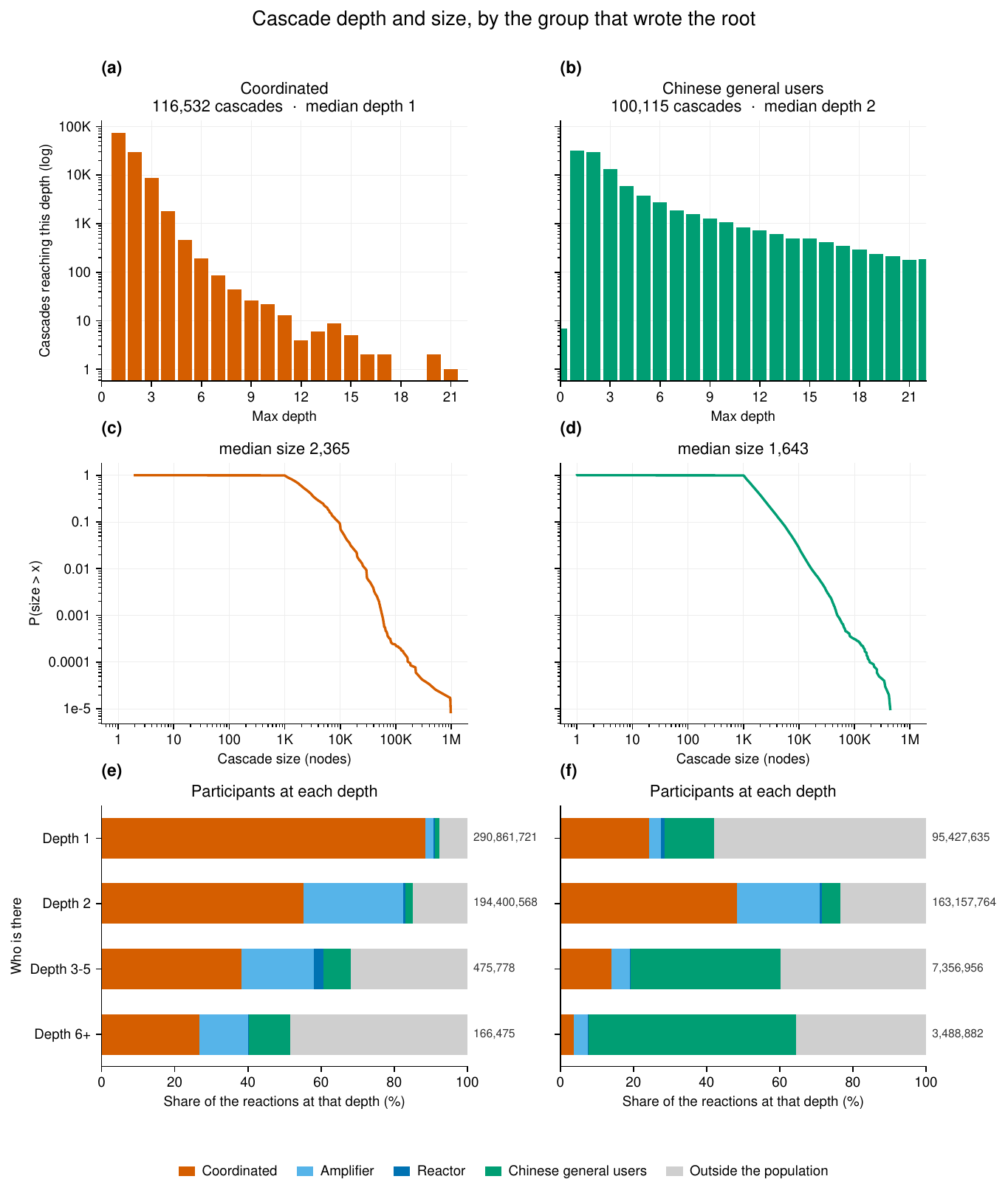}
\caption{Cascade distributions by the group of the account that wrote the root: coordinated accounts (a, c, e) and Chinese general users (b, d, f). (a--b) Maximum depth; (c--d) cascade size; (e--f) share of reactions at each depth by the group of the reacting account.}
\label{fig:cascade-dist}
\end{figure}

Figure~\ref{fig:cascade-dist} compares the maximum depth and size of cascades generated by different account types. It contrasts cascades rooted in posts by coordinated accounts with those originating from general Chinese accounts.


\subsection{Who shares at each depth} 
\begin{figure*}[!t]
\centering
\includegraphics[width=1.1\textwidth]{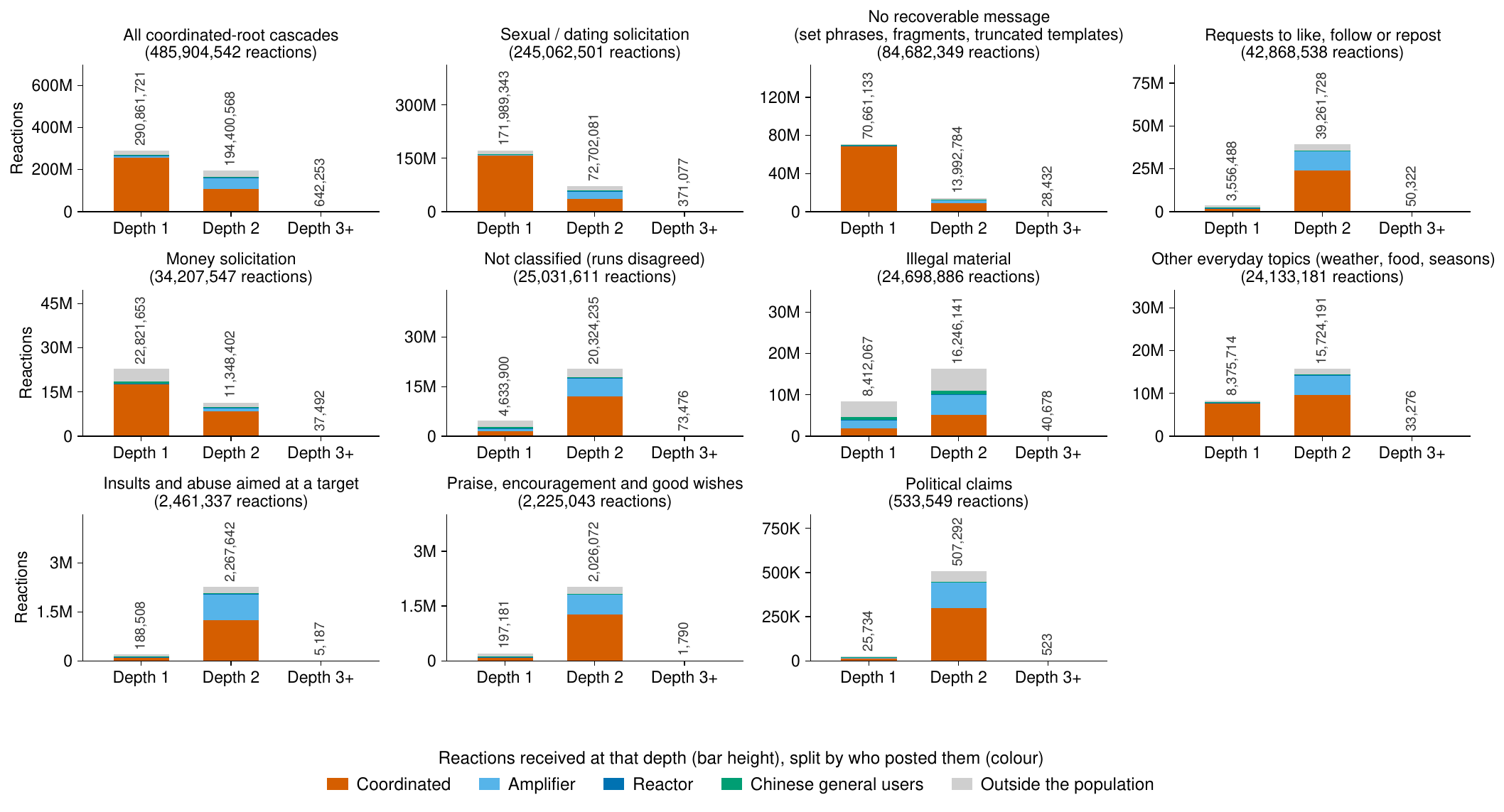}
\caption{Who reacted at each depth, by the content of the root post}
\label{fig:cascade-topic}
\end{figure*}

Figure~\ref{fig:cascade-topic} decomposes the cascades by depth within each root category. We report it here rather than in the main text because the deep bars rest on quite few nodes: at depth five and beyond, seven of the eleven categories have fewer than 25 nodes in this sample. Node counts are printed above every bar so that no bar is read without its n.

\subsection{Most frequent coordinated texts}

Table~\ref{tab:top-coord} lists the thirty most frequent normalized texts posted by coordinated accounts, with an English gloss, the total number of posts, and the most frequent unnormalized form. 

\begin{table*}[htbp]
\centering
\includegraphics[width=0.95\textwidth]{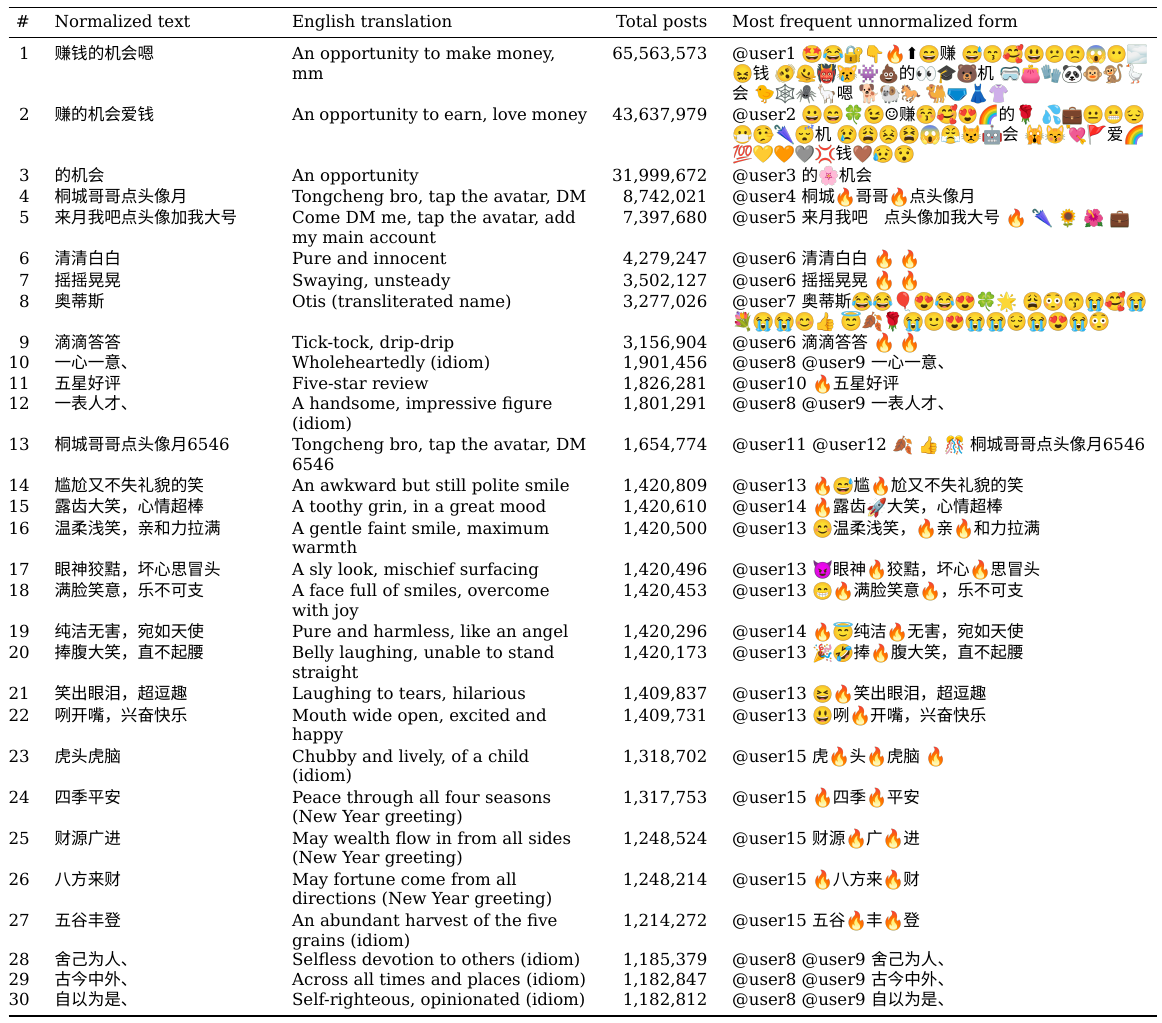}
\caption{The 30 most frequent texts posted by coordinated accounts, after normalization}
\label{tab:top-coord}
\end{table*}

\subsection{Number of suspended accounts}

By utilizing the X API, we found that 710,595 out of the 734,173 coordinated accounts (96.79\%) had been suspended by X at the time of our query. Across all topics, at least 84\% of the coordinated accounts were suspended.

\begin{figure*}[t]
\centering
\includegraphics[width=\textwidth]{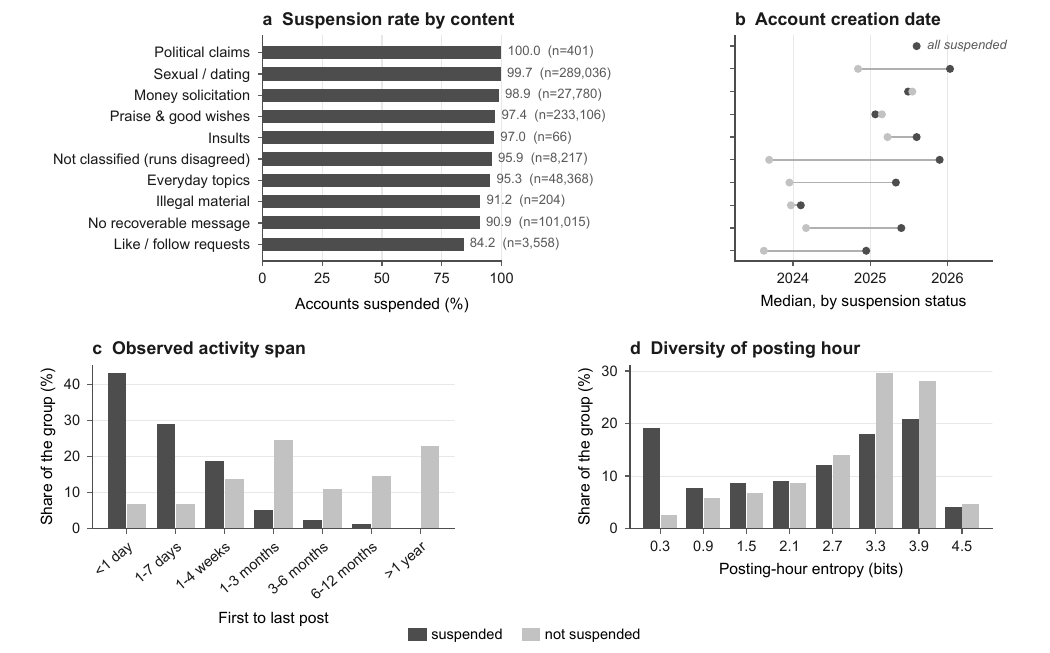}
\caption{Suspension rates by content category (a) and characteristics of suspended versus non-suspended accounts (b--d)}
\label{fig:suspension}
\end{figure*}

\subsection{Reach by vocabulary rather than by category}
\label{app:vocab-reach}

Because human validation revealed boundaries in the illegal-material category to be imprecise, we partitioned coordinated-root cascades by vocabulary rather than by assigned category. Roots were matched against three predefined term lists: adult sexual terms, terms denoting minors, and terms related to the illicit trade of personal data. Within the illegal-material category, roots containing terms that denote minors ($n = 824$) reached non-coordinated accounts at a rate of 44.1\%, while those containing adult sexual terms ($n = 3{,}684$) reached 41.2\%. Similarly, within the sexual solicitation category, roots carrying adult sexual terms ($n = 9{,}994$) achieved an external reach of 36.2\%, compared to 10.7\% for the category overall ($n = 69{,}029$). The reach of a coordinated post thus tracks its specific vocabulary rather than its assigned classification. Two limitations apply: this partitioning relies on term matching rather than manual re-annotation, and 2,299 roots (33.8\%) in the illegal-material category matched none of the three lists.

\subsection{A baseline outside the Chinese-language population}
\label{app:baseline}
The share of accounts meeting our coordination criterion is only interpretable against some baseline. We therefore applied the identical procedure---the same kana split, the same minimum of thirteen body posts, the same normalization, and the same thresholds of three distinct accounts and three repetitions---to both populations over a single month, March 2025, which lies between the two surges and so reflects neither population at peak. The month was fixed because the criterion counts repetitions within a period. Under these conditions 58,331 of 92,548 Chinese-language accounts (63.0\%) and 61,206 of 573,058 Japanese-language accounts (10.7\%) met the criterion. The Japanese-language population is larger and posts more in absolute terms (42.8 million body posts against 11.6 million), so the difference is not an artifact of corpus size.

\end{document}